\documentclass[preprint,review,12pt]{elsarticle}

\usepackage{amssymb}
\journal{Physics Letters A}

\begin{document}

\begin{frontmatter}



\title{Consequences of an attractive force on collective modes and dust structures in a strongly coupled dusty plasma}
\author[label1]{Z. Donk\'o\corref{cor1}}
\ead{donko.zoltan@wigner.mta.hu}
\cortext[cor1]{Corresponding author. Tel.: +36 1 392 2222; fax: +36 1 392 2215}

\author[label1]{P. Hartmann}

\address[label1]{Institute of Solid State Physics and Optics, Wigner Research Centre for Physics, Hungarian Academy of  Sciences, \\ H-1121 Budapest, Konkoly Thege M. str. 29-33, Hungary}

\author[label2]{P. K. Shukla}
\address[label2]{International Centre for Advanced Studies in Physical Sciences \& Institute for Theoretical Physics, Faculty of Physics \& Astronomie, Ruhr University Bochum, D-44780 Bochum, Germany, and Department of Mechanical and Aerospace Engineering \& Center for Energy Research, University of California San Diego,  La Jolla, CA 92093, U. S. A}

\begin{abstract}
We present investigations of the combined effects of Debye-H\"uckel repulsive and overlaping Debye spheres attractive interaction potentials around charged dust particles on collective modes, phase separation and ordered structures in a strongly coupled dusty plasma. We obtain static and dynamical information via Molecular Dynamics simulations in the liquid and crystallized phases and identify the  onset of an instability in the transverse mode, by using lattice summation method. The results are useful for understanding the origin of coagulation/agglomeration of charged dust particles and the formation of ordered dust structures in low-temperature laboratory and space plasmas.
\end{abstract}

\begin{keyword}
dusty plasmas \sep attractive force \sep molecular dynamics simulation \sep collective excitations
\end{keyword}

\end{frontmatter}

\section{Introduction}
\label{intro}

Dusty plasmas are composed of electrons, ions and dust particles of different sizes (ranging from nanometer to micrometer). Dust particles, which are negatively charged by absorbing electrons from the background plasma, respond to electrostatic and electromagnetic fields and become active components 
of the dusty plasma system, especially when their dynamics comes into the picture at kinetic levels \cite{SM2002,Fortov2010}. Dusty plasmas are ubiquitous in many physical environments, viz. in interstellar media and nebulas, in our solar sysem (e.g. Saturn's ring system, cometary tails and comas, in in the Earth's mesosphere and ionosphere), in industrial and laboratory processing plasmas for high-technological applications (also in medicine), as well  in laboratory gas discharges and  magnetically confined fusion reactors. In low-temperature  laboratory dusty plasmas on ground and on International Space Station under microgravity,  dust is intentionally introduced  in plasmas for fundamental studies of collective phenomena at kinetic  level (e.g. the discovey of the dust acoustic wave \cite{RSY1990}), involving an ensemble of charged dust grains.  In astrophysics, the coagulation of charged dust particulates plays a significant role in the formation of large scale structures (e.g. plenetesimals). There are forthcoming space missions for exploring the composition and dynamics of dust particles in the vicinity of Sun,  as well on Mars and Moon.

To describe the interaction between the charged  dust particles in dusty plasma systems, the Debye-H\"uckel (DH) potential $\phi^{DH} (r)$ has been widely adopted as a model potential around a chargd dust grain
that has a well defined surface potential. The DH potential expresses that the Coulomb repulsion between 
the charged gains of similar polarity is screened by the plasma environment, where electrons and ions represent a polarizable background. Consequently, the $1/r$ Coulomb potential is exponentially shielded over a characteristics distance, the dusty plasma Debye radius \cite{SM2002} $\lambda_D$, and we have  

\begin{equation}
\phi^{DH}(r) = \frac{Q}{4 \pi \varepsilon_0} \frac {\exp(-r/\lambda_{\rm D})} {r} ,
\label{eq:dhpot}
\end{equation}
where $Q$ is the charge of the dust particles and $\varepsilon_0$ is the permittivity of free space.

At short inter-particle distances, the DH interaction potential has to be modified, since the Debye spheres of the neighboring grains would overlap \cite{RMS1998}. From classical electrostatics, one can then derive an additional term of the interparticle potential that expresses the interaction between these overlapping Debye spheres (ODS) \cite{RMS1998,SE2009}. The resulting (total) potential in this case becomes

\begin{equation}
  \phi^{ODS}(r) = \frac{Q}{4 \pi \varepsilon_0}  
  \Bigl[ \frac {1} {r} - \frac {1} {2 \lambda_D}  \Bigr] 
  ~\exp(-r/\lambda_{\rm D}),
\label{eq:odspot}
\end{equation}
which exhibits an attractive force beteen charged dust grains of the same polarity at distances beyond 
$r = \sqrt{3} \lambda_D$. 

In this Letter, we investigate the effect of this peculiarity of the attractive potential on the structure of two-dimensional dust-plasma systems in the crystallized phase and in the strongly coupled liquid phase, 
where the potential energy dominates over the kinetic energy, i.e. $\Gamma \gg 1$, where $\Gamma$
 is the coupling parameter, viz.

\begin{equation}
  \Gamma = \frac{Q^2} {4 \pi \varepsilon_0 a k_B T}.
\label{eq:gamma}
\end{equation}
Here $k_B$ is the Boltzmann constant, $T$  the dust temperature, and $a = (1 /  \pi n)^{1/2}$ the Wigner-Seitz radius (that characterizes the inter-dust particle spacing), with $n$ being the dust particle number density. We also introduce the dimensionless screening parameter

\begin{equation}
  \kappa = \frac{a}{\lambda_D}.
\label{eq:kappa}
\end{equation}

In a previous theoreical investigation \cite{Chang2005}, the attractive term in $\phi^{ODS}(r)$ has been shown  to cause instability of transverse dust lattice oscillations. Considering the nearest neighbor interactions only, the onset of  instability was identified at $b / \lambda_D > \sqrt{3} +1$, where $b$ is the lattice constant. For the triangular lattice $b \approx 1.905 a$, the critical screening value in terms of the inter-grain spacing becomes $\kappa_c = a / \lambda_D \approx$ 1.43. 

As regards to the strongly coupled liquid phase, Molecular Dynamics simulations have confirmed that $\phi^{ODS}(r)$ gives rise to an uneven spatial distribution (agglomeration) of dust particles \cite{Hou2009}. 

\section{Methods}

In our numerical work, we use a standard Molecular Dynamics simulation method with Langevin dynamics \cite{Hou1,Hou2}. The system is two-dimensional and we use periodic boundary conditions. We carry out computations with both potentials, $\phi^{DH}(r)$ and $\phi^{ODS}(r)$. The simulations are initialized either by a spatially random particle configuration (in the studies of the liquid phase, with $N$ = 10,000 dust particles) or by a lattice configuration (in the studies of the crystallized phase, with $N$ = 9180 dust particles). The side lengths of the simulation box are chosen to accommodate a perfect triangular lattice. Initial dust particle velocities are sampled from a Maxwellian distribution of the temperature $T_0$, corresponding to the specified value of the coupling parameter $\Gamma$. The interaction between the dust particles is limited to distances smaller than a cutoff radius, $r^\ast$, which is chosen to be large enough so that interparticle forces at $r > r^\ast$ become very small due to the exponential term in the interaction potential.  The neighbors of the dust particles within $r < r^\ast$ are searched using the chaining mesh technique. 

The equation of motion includes terms representing the frictional damping and random forces, which express random collisions of gas molecules with the charged dust particles. We have

\begin{equation}
  m\frac{d {\bf v}_i}{dt}=-Q \sum_{j \neq i}^N \nabla \phi_{ij} - \nu m {\bf v} +{\bf R}, \\
\label{eq:lmotion}
\end{equation}
where $\phi_{ij}$ is the interaction potential between dust particles $i$ and $j$, $\nu$ is the frictional damping, and ${\bf R}$ is a random force, see e.g. \cite{Hou1,Hou2}. In all our simulations, we use a small friction value, $\Theta = \nu / \omega_0 = 0.01$, where $\omega_0 = \sqrt{n Q^2 / 2 \varepsilon_0 m a}$ is the nominal 2D dust plasma frequency, with $m$ being the dust particle mass. We introduce dimensionless distance as $\bar{r} = r/a$ and the wave number as $\bar{k}=ka$.

The pair correlation function and the static structure function are derived in the simulations in the standard manner. Information about the (thermally excited) collective modes and their dispersion is obtained from the Fourier analysis of the correlation spectra of the density fluctuations,

\begin{equation}\label{eq:rho}
\rho(k,t)= \sum_j \exp \bigl[ i k x_j(t) \bigr],
\end{equation}
yielding the dynamical structure function as \cite{HMP75}:

\begin{equation}\label{eq:sp1}
S(k,\omega) = \frac{1}{2 \pi N} \lim_{\Delta t \rightarrow \infty}
\frac{1}{\Delta t} | \rho(k,\omega) |^2,
\end{equation}
where $\Delta t$ is the length of data recording period and $\rho(k,\omega) = {\cal{F}} \bigl[ \rho(k,t) \bigr]$ is the Fourier transform of (\ref{eq:rho}).

Similarly, the spectra of the longitudinal and transverse current fluctuations, $L(k,\omega)$ and $T(k,\omega)$ can be obtained from Fourier analysis of the microscopic quantities, respectively,

\begin{eqnarray}
\lambda(k,t)= \sum_j v_{j x}(t) \exp \bigl[ i k x_j(t) \bigr], \nonumber \\
\tau(k,t)= \sum_j v_{j y}(t) \exp \bigl[ i k x_j(t) \bigr],
\label{eq:dyn}
\end{eqnarray}
where $x_j$ and $v_j$ are the position and velocity of the $j$-th particle. Here we assume that ${\bf k}$ is directed along the $x$ axis (the system is isotropic) and accordingly omit the vector notation of the wave number. This calculation allows the derivation of the spectra for a series of wave numbers, which are multiples of $k_{min,x(y)} = 2 \pi / L_{x(y)}$, where $L_{x(y)}$ is the edge length of the simulation box 
in the $x$ (or $y$) direction. The collective modes are identified as peaks in the fluctuation spectra. 

The phonon dispersion in the lattice configuration is calculated in terms of the lattice dynamical matrix defined as

\begin{eqnarray}
C_{\mu\nu}({\bf k})=-\frac{Q}{m}\sum_{i,j}\partial_{\mu}\partial_{\nu} \phi(r)\left( {\bf
r}_i-{\bf r}_j\right) \times \\ \nonumber
 \left[ {\rm e}^{-i{\bf k}\cdot ( {\bf r}_i-{\bf r}_j)}-1\right],
\end{eqnarray}
with a summation over all the lattice sites $j$, keeping $i$ fixed $({\bf r}_i=0)$. The lattice
dynamical matrix reflects the symmetry of the underlying lattice. The diagonalization of $C_{\mu\nu}$ is possible in the coordinate system of the eigenvectors, whose orientations, in general, do not coincide 
either with the direction of ${\bf k}$ or with the crystallographic axes. To find the eigenmodes, we follow
the traditional method of solving the secular equation

\begin{equation}
||\omega^2-C_{\mu\nu}({\bf k})||=0.
\end{equation}

\section{Results}
\label{sec:results}

\begin{figure*}
\resizebox{1\columnwidth}{!}{\includegraphics{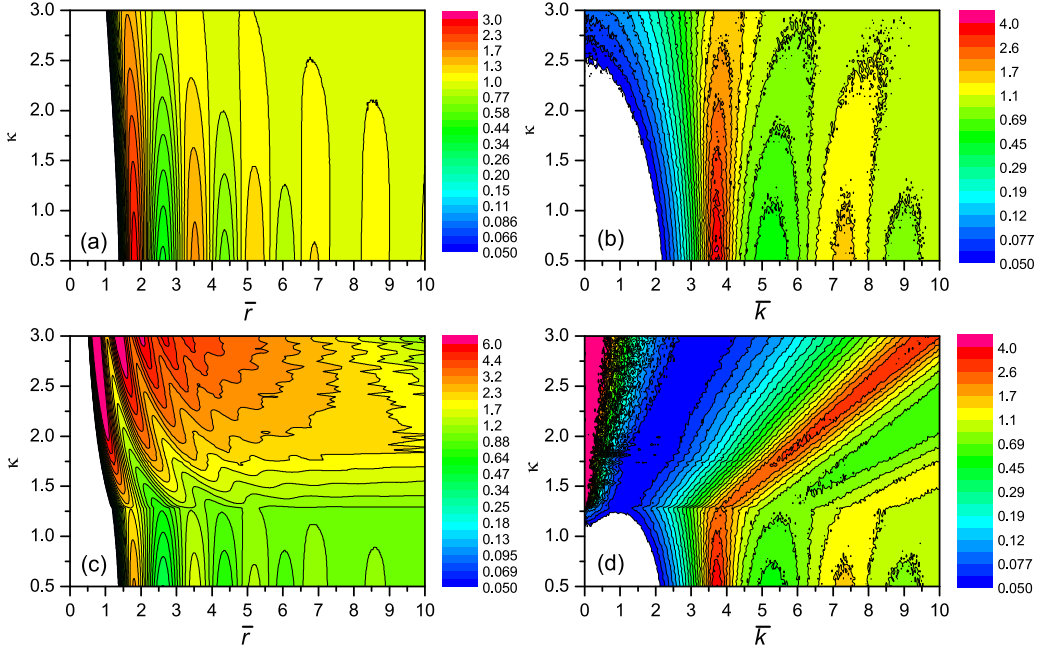}}
\caption{Static characteristics of the 2D dusty plasma obtained with $\phi^{DH}$ (top row) and with $\phi^{ODS}$ (bottom row). (a,c) pair correlation function, $g(r/a)$, (b,d) static structure function $S(k)$. $\Gamma = 120.$}
\label{fig:static}       
\end{figure*}

In figure \ref{fig:static} we present an overview of the main static characteristics of our dusty plasma system: the four panels display the pair correlation function [$g(r)$] and the static structure function [$S(k)$] for the $0.5 \leq \kappa \leq 3$ domain of the screening parameter, for both types of potentials. The coupling parameter is fixed at $\Gamma$= 120; at this value the system is in the liquid phase for the whole range 
of $\kappa$ values covered. Panels (a) and (b) correspond to $\phi^{DH}$, while (c) and (d) correspond to $\phi^{ODS}$. At small screening values, viz. $\kappa \lesssim 1$, the $g(r)$ and $S(k)$ functions are similar for both potentials, showing slightly less organized structures (indicated by smaller amplitude the peaks) in the case of $\phi^{ODS}$. In the case of $\phi^{ODS}$, we observe major changes in both static quantities in the vicinity of the screening value $\kappa \approx$ 1.3. In $g(r)$ a strong peak develops at small separations, indicating particle agglomeration and $S(k)$ suddenly increases at small wave numbers, indicating development of large-scale structures.   

\begin{figure*}
\resizebox{1\columnwidth}{!}{\includegraphics{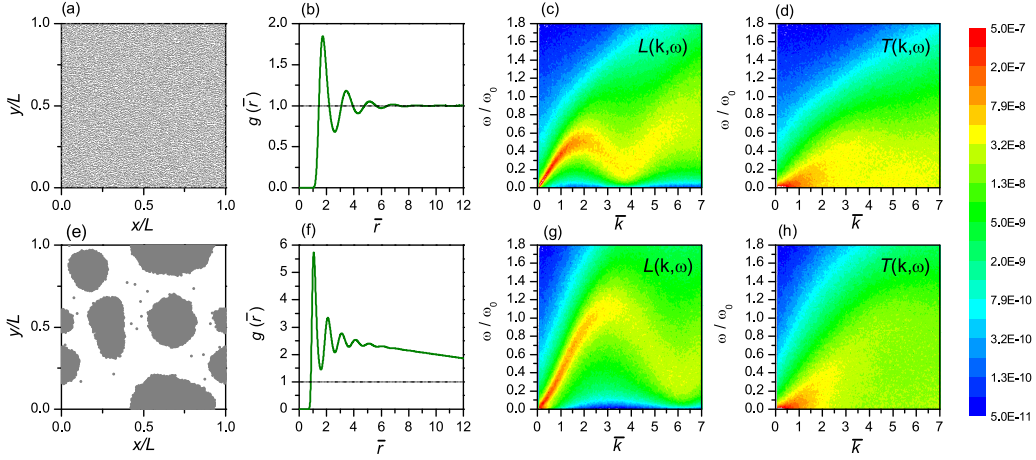}}
\caption{MD simulation results obtained with the potentials $\phi^{DH}$ (top row) and $\phi^{ODS}$ (bottom row). (a,e) Snapshots of particle positions, (b,f) pair correlation functions, (c,g) spectra of longitudinal current fluctuations, and (d,h) spectra of transverse current fluctuations. $\Gamma$ = 100, $\kappa = 2$.}
\label{fig:comp}       
\end{figure*}

\begin{figure*}
\resizebox{1\columnwidth}{!}{\includegraphics{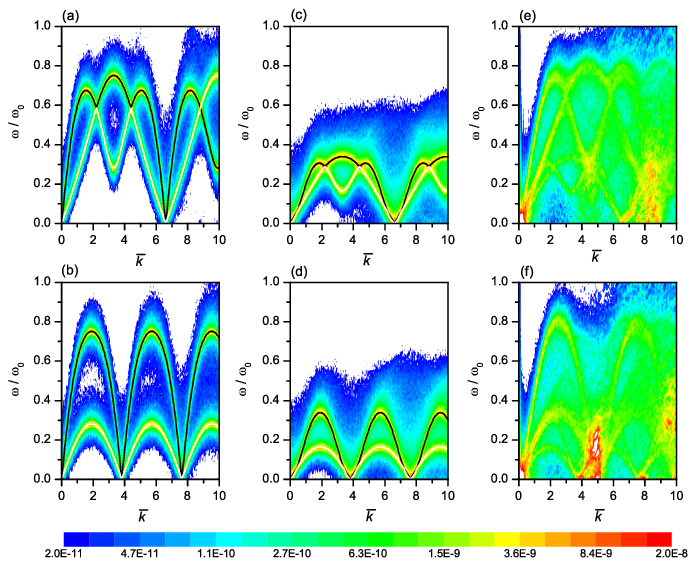}}
\caption{Sum of the current fluctuation spectra, $L(k,\omega)+T(k,\omega)$, derived from the MD simulations (color maps) and dispersion relations of the collective excitations obtained from zero-temperature lattice summation (solid lines: black: longitudinal modes, white: transverse modes). Upper row (a,c,e): propagation direction: $\alpha = 0^\circ$, bottom row (b,d,f): $\alpha = 30^\circ$.  (a,b) $\phi^{DH}$ with $\kappa$ = 1, (c,d) $\phi^{ODS}$ with $\kappa$ = 1.47, (e,f) $\phi^{ODS}$ with $\kappa$ = 1.48. All MD runs have been initialized with lattice configuration of particles, at $\Gamma = 10^4$. 
For the color maps the scale is logarithmic and covers three orders of magnitude.}
\label{fig:latt}       
\end{figure*}

Figure \ref{fig:comp} shows a comparison between systems characterized by $\phi^{DH}$ and $\phi^{ODS}$, at $\Gamma=100$ and $\kappa=2$. At this screening strength, the system with $\phi^{ODS}$ already exhibits agglomeration, as the particle snapshot and the corresponding pair correlation function (in panels (e) and (f) of figure \ref{fig:comp}) illustrate. A notable difference in the $L(k,\omega)$ spectrum is the significant increase of the maximum mode frequency from $\omega/\omega_0 \approx 0.5$ (at $\phi^{DH}$) to $\approx $1.1 (at $\phi^{ODS}$), and an increase of the wave number where this frequency is found, from $\overline{k} \approx $ 1.9 to $\approx$ 3.1. As the transverse mode is quite weak in the liquid phase, it is more difficult to characterize quantitatively the frequency of this mode, but an increase of the frequency is observable, too. 

Next, we compare the results of our MD simulations with those obtained from lattice summation. To be able to do that, we have initialized the simulations with a lattice configuration and set a very high coupling value, $\Gamma = 10^4$. Figure \ref{fig:latt} shows the results for three cases: (i) $\phi^{DH}$ with $\kappa=1$, (ii) $\phi^{ODS}$ with $\kappa=1.47$, and (iii) $\phi^{ODS}$ with $\kappa=1.48$ The lattice is stable in the first two of these three cases, while the third case belongs to the unstable domain with $\phi^{ODS}$. The MD and lattice calculation results are in very good agreement for the stable cases. The figure exhibits that 
the L and T modes are superimposed, at propagation directions of $\alpha = 0^\circ$ (the direction of the nearest neighbor) and $\alpha = 30^\circ$. In these propagation directions, the polarization of the modes becomes purely longitudinal and transverse, in other directions the polarizations mix. At $\kappa = 1.49$ complicated structures  appear in the fluctuation spectra that are difficult to interpret.

\begin{figure*}
\resizebox{1\columnwidth}{!}{\includegraphics{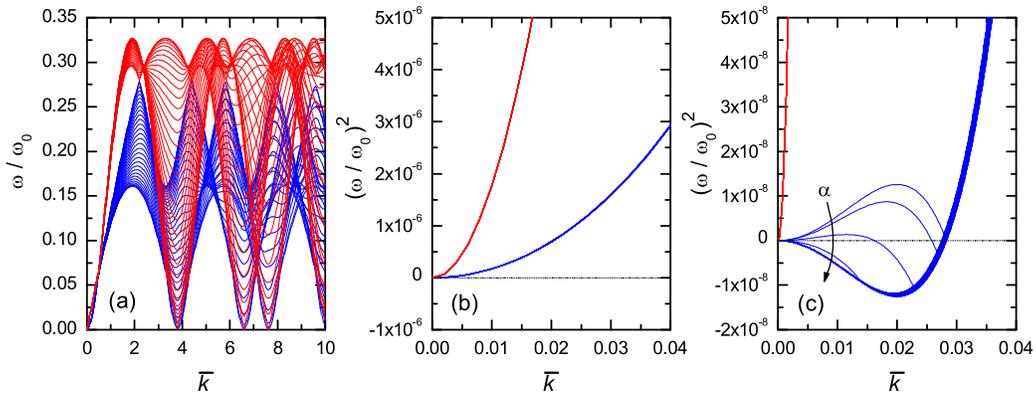}}
\caption{Results of lattice summations for the eigenfrequencies of the two modes, for $\phi^{ODS}$. Red color denotes quasi-longitudinal and blue color denotes quasi-transverse modes; note that the polarization of the modes is purely longitudinal or transverse in the principal directions ($\alpha = 0^\circ$ and $\alpha = 30^\circ$) only. The panels show the frequencies in propagation directions between $\alpha = 0^\circ$ and $\alpha = 30^\circ$ with $1^\circ$ resolution. (a) Stable domain, $\kappa = 1.49$.  (b) Zoomed part of (a) showing the square of the mode frequencies at small wave numbers. In this domain, all dispersion curves for $\alpha = 0^\circ \dots 30^\circ$ overlap. (c) Same as (b), but for $\kappa = 1.50$; note the splitting of the dispersion curves and the appearance of negative $(\omega/\omega_0)^2$ values, indicative of an instability. The arrow shows increasing values of $\alpha$.}
\label{fig:disp}       
\end{figure*}

Figure \ref{fig:disp} displays the results of lattice summations for the eigenfrequencies of the two modes, for $\phi^{ODS}$, characterizing the interaction of dust particles with overlapping Debye spheres. Figure \ref{fig:disp}(a) shows the dispersion relation of the modes with different propagation directions, for 
$\kappa = 1.49$, which belongs to the stable regime of the lattice at $T$ = 0. The two branches of the curves correspond to ``quasi-longitudinal'' and ``quasi-transverse'' modes. The polarization of the modes 
is usually mixed, except for the principal directions ($\alpha = 0^\circ$ and $\alpha = 30^\circ$), where the polarization is purely longitudinal or transverse. Figure \ref{fig:disp}(b) exhibits $(\omega/\omega_0)^2$ values for small wave numbers at $\kappa = 1.49$. In this domain, all dispersion curves for $\alpha = 0^\circ \dots 30^\circ$ overlap. Figure \ref{fig:disp}(c) shows also a zoomed part of the mode dispersion relations, but for $\kappa = 1.50$; at this screening value negative $(\omega/\omega_0)^2$ values appear that correspond to an instability of the lattice. 

The critical value of the screening parameter was found to be $\kappa_c  = 1.4997$ in our lattice calculations (that add contributions to the dynamical matrix up to the 126th neighbors of the particles). The theoretical calculations \cite{Chang2005} taking into account the nearest neighbor interactions only resulted in $\kappa_c \approx$ 1.43.  In our MD simulations carried out for $\Gamma = 10^4$ we found that the lattice disintegrates between $\kappa = 1.47$ and $\kappa = 1.48$, while in the fluid phase, as shown in Figure \ref{fig:static} for $\Gamma=120$, an instability occurs at $\kappa_c  \approx$ 1.3. These changes indicate the effect of the temperature on the development of the instability.

\section{Summary} 

Summing up, we have investigated the effect of an attractive term of the interaction potential in addition to 
the DH potential, conventionally used to model dusty plasmas, on the static and dynamical characteristics 
of many-body dust-plasma systems. Molecular dynamics simulations (with a Langevin approach) carried out in the liquid phase showed that the onset of particle agglomeration at screening values exceeding a critical value. The particle agglomeration was accompanied with a significant increase of the frequencies of the longitudinal and transverse collective excitations. In the crystallized phase, the MD simulations revealed that the hexagonal lattice becomes unstable at high screening. These observations have been confirmed by harmonic lattice summations in the zero temperature limit, which showed the development of an instability 
in the (quasi-)transverse modes at the critical screening value of $\kappa_c  \approx 1.4997$. In conclusion, we stress that the results of the present investigation are useful for understanding the origin of agglomaration/coagulation of charged dust particles and  the formation of ordered dust structures 
in low-temparature laboratory and space dusty plasmas \cite{Fortov2010}.
 
This work was supported by the Hungarian Fund for Scientific Research through grants K77653, K105476, NN103150, and by the HELIOS project (ELI 09-1-2010-0010).

\end{document}